\documentclass[twocolumn,floatfix,aps,prc,showpacs]{revtex4}
 
\usepackage[final]{epsfig}
\usepackage{amsmath}
\usepackage{bm}
\usepackage{graphicx}

\usepackage{color}

\begin{document}
 
\title{Constraining the Physics of Jet Quenching}
 
\author{Thorsten Renk}
\email{thorsten.i.renk@jyu.fi}

\affiliation{Department of Physics, P.O. Box 35, FI-40014 University of 
             Jyv\"askyl\"a, Finland}
\affiliation{Helsinki Institute of Physics, P.O. Box 64, FI-00014 University 
             of Helsinki, Finland}


\begin{abstract}
Hard probes in the context of ultrarelativistic heavy ion collisions represent a key class of observables studied to gain informations about the QCD medium created in such collisions. However, in practice the so-called jet tomography has turned out to be more difficult than expected initially. One of the major obstacles in extracting reliable tomographic information from the data is that neither the parton-medium interaction nor the medium geometry are known with great precision, and thus a difference in model assumptions in the hard perturbative Quantum Choromdynamics (pQCD) modelling can usually be compensated by a corresponding change of assumptions in the soft bulk medium sector and vice versa. The only way to overcome this problem is to study the full systematics of combinations of parton-medium interaction and bulk medium evolution models. This work presents a meta-analysis summarizing results from a number of such systematical studies and discusses in detail how certain data sets provide specific constraints for models. Combining all available information, only a small group of models exhibiting certain characteristic features consistent with a pQCD picture of parton-medium interaction is found to be viable given the data. In this picture, the dominant mechanism is medium-induced radiation combined with a surprisingly small component of elastic energy transfer into the medium.

\end{abstract}
 
\date{\today}

\maketitle

\section{Introduction}

Hard probes, i.e. high transverse momentum ($P_T$) processes in the context of heavy ion have long been regarded as a useful tool to gain tomographic information about the medium produced in ultrarelativistic heavy-ion collisions \cite{radiative1,radiative2,radiative3,radiative4,radiative5,radiative6}. The idea underlying tomography is that high $p_T$ Quantum Chromodynamics (QCD) processes can be reliably calculated using factorized perturbative QCD (pQCD) in vacuum, and since the uncertainty relation ensures that the scales of medium physics and hard process are well separated, this implies that high $p_T$ parton production takes place in a heavy-ion (A-A) collision with a calculable rate, which is approximately equal to the vacuum rate corrected for the number of binary nucleon-nucleon collisions in an A-A collision. The effect of the medium is then  that of a final state interaction, i.e. chiefly a modification of the outgoing parton fragmentation pattern due to parton-medium interactions. As in the case of X-ray tomography, having a known process embedded into a medium can then in principle be used to image this medium and obtain information about its density distribution.

There are, however, notable differences to X-ray tomography which have prevented this idea from being fully utilized so far. First of all, in X-ray tomography the interaction between photon and medium is known to a great precision, whereas in high $P_T$ tomography the parton-medium interaction depends even qualitatively on the \emph{a priori} unknown degrees of freedom of the medium as probed at the resolution scale of the propagating parton. Second, while in X-ray tomography a monochromatic photon source with a known position is utilized, in high $P_T$ tomography both primary parton momentum and position are only known probabilistically. Third, while the medium in X-ray tomography is effectively static at the timescale of photon propagation, in heavy-ion collisions it undergoes a violent expansion at speeds of a sizable fraction of the speed of light $c$ and hence changes during the parton propagation.

The combination of these effects implies that due to the necessary averaging over probabilistically known initial variables the sensitivity of observables to the desired properties of the medium, i.e. its relevant degrees of freedom and its density evolution, is not as large as one would naively expect \cite{Deconvolution}, while at the same time there is an ambiguity between model assumption made with regard to the bulk density evolution and with regard to the parton-medium interaction physics. This is exemplified e.g. by the work in \cite{PHENIX-constraints} --- while a detailed comparison of models with the data is able to constrain one relevant model parameter with good precision assuming an otherwise perfect theory, there is no consensus between different models as to what this implies for the parton-medium interaction physics or the medium density evolution. 

The only way to overcome these obstacles is to study the full systematics of both medium and parton-medium interaction modelling with a sufficiently large body of data. While certain parts of the problem have been investigated in various papers, a full overview about what is known and what constraints the combined high $P_T$ data provide has not been available so far. It is the aim of this work to present such a meta-analysis of several systematic studies and to argue that by making use of the large body of available data, the features of parton-medium interaction and to a lesser degree also the density evolution of the bulk medium can be reasonably well constrained.

This work is organized as follows: First, an overview over experimentally available high $P_T$ observables and their characteristics is given along with arguments justifying the selection of observables discussed in this study. In section \ref{S-Assumptions}, an overview over the different classes of assumptions which have to be made when specifying a model for hard probes is given, as well with a review of what is known about the sensitivity of certain observables to these assumptions. Section \ref{S-Constraints} then summarized what constraints can be derived from selected observables using a systematic analysis of a large body of models. Section \ref{S-OpenIssues} proceeds by providing supplementary information on what is known about several other observables and other types of models where currently no systematic results are available and how these results relate to the previous section. Finally, a summary of the essentail results is provided.

\section{Experimental observables}

High $P_T$ tomography utilizes a large number of observables raning from single inclusive hadron suppression to fully reconstructed jets. Each of these observables has its own advantages and disadvantages, and the following list aims to illustrate the gross characteristics, strength and weaknesses of each and introduces the relevant terminology.

\subsection{Single inclusive hadron suppression}

The least involved observable is the single inclusive hadron suppression factor $R_{AA}$ which is defined as the yield of high $P_T$ hadrons from an A-A collision normalized to the yield in p-p collisions at the same energy corrected for the number of binary collisions, 

\begin{equation}
\label{E-RAA}
R_{AA}(p_T,y) = \frac{dN^h_{AA}/dp_Tdy }{T_{AA}({\bf b}) d\sigma^{pp}/dp_Tdy}.
\end{equation}

The normalization with the number of binary collisions ensures $R_{AA} = 1$ in the absence of non-trivial initial or final state nuclear effects. The default expectation is $R_{AA} < 1$ in medium since parton-medium interaction is expected to lead to a flow of high $p_T$ parton momentum into medium degrees of freedom, thus effectively suppressing the yield in any given momentum bin. In the absence of final state effects, nuclear initial state effects can however cause $R_{AA} > 1$ in some kinematical regions \cite{KinematicLimit}.

Experimentally, $R_{AA}$ can readily be obtained for different colliding ion species, as a function of collision centrality, as a function of $\sqrt{s}$ and, if a reaction plane (or $v_n$ event plane where $v_n$ is the $n$th coefficient in a harmonic expansion) is identified, also for given centrality as a function of the angle $\phi$ of high $P_T$ hadrons with the reaction plane ($v_n$ event plane). 

If $R_{AA}$ is obtained as a function of $\phi$, then the spread $S^{in}_{out}  = R_{AA}(0) - R_{AA}(\pi/2)$ between in plane and out of plane emission is an important observable sensitive to the medium geometry. Knowledge of $R_{AA}(0)$  and $R_{AA}(\pi/2)$ is exactly equivalent to angular averaged $R_{AA}$ and the second harmonic coefficient $v_2$ at high $P_T$, however a discussion in terms of $R_{AA}$ emphasizes that unlike at low $P_T$ the asymmetry in momentum space is created by an extinction process rather than by pressure gradients as at low $P_T$, therefore we will not discuss high $P_T$ phenomena in terms of $v_2$ in the following.

\subsection{Dihadron correlation suppression}

The direct observables in dihadron correlation measurements are conditional yields, the so-called yield per trigger (YPT) on the near and away side. These depend crucially on the momentum range required for a trigger hadron. The YPT do not directly reflect any suppression of the observed rate of triggered events in A-A collisions compared to a p-p reference (which is closely related to the nuclear suppression factor $R_{AA}$). Thus, for instance in an initial state suppression picture where a back-to-back parton event is either suppressed or never feels a medium, $R_{AA}$ can be at an arbitrarily low value while the  YPT is unchanged in the medium.

The near and away side YPT is often binned in momentum windows, but sometimes also in terms of the fraction of the associate hadron momentum divided by trigger hadron momentum, $z_T = P_T^{assoc}/P_T^{trig}$. Since $z_T$ has a probabilistic connection to the fractional momentum $z = P_T^{had}/p_T^{part}$ of a hadron produced from a parton with momentum $p_T^{part}$, the away side distribution binned in terms of $z_T$ is often called $D(z_T)$ similar to the fragmentation function. 

When comparing the results from A-A and p-p collisions, usually the yield ratios $I_{AA} = Y_{med}/Y_{vac}$ (where $Y_{med}$ and $Y_{vac}$ are the YPT in medium and vacuum respectively) are formed where $I_{AA} = 1$ indicates the absence of a medium modification to the correlation strength. Unlike in the case of $R_{AA}$, there is no strong {\it a priori} expectation $I_{AA} < 1$ in the medium. Various biases (discussed below) are potentially capable of generating $I_{AA} >1$ for $R_{AA} < 1$ under the right kinematical conditions.

\subsection{$\gamma$-hadron correlations}

If one requires the trigger particle to be a photon and correlates any associated away side yield, the resulting observable is conceptually more powerful than the dihadron correlations discussed above. In $\gamma$-hadron correlations, photons produced in the hard process itself undergo neither fragmentation nor final state interactions with the medium, and as a result they carry almost undistorted information about the original kinematics at the hard vertex, i.e. $z_T \approx z$ and $D(z_T) \approx D(z)$ for a photon trigger. These relations are exactly true in leading order calculations but not at higher orders where hard gluon radiation can introduce a momentum imbalance. Moreover, a photon may also be produced in a parton shower rather than directly at the vertex, which further dilutes the momentum information.

Since the dominant leading order production channel is the QCD compton scattering $qg \rightarrow q \gamma$ and the anihilation process $q\overline{q} \rightarrow g \gamma$ is combinatorically suppressed, to a good approximation the away side parton given a photon trigger is a quark. This allows to specifically study the quark-medium interaction.

An obvious experimental disadvantage is that due to the smallness of the electromagnetic coupling $\alpha_{em}$ relative to the strong coupling constant $\alpha_s$ and the need to identify photons from the hard vertex out of a large number of other photon sources the statistics of $\gamma$-hadron correlation is usually much worse than for dihadron correlations.

\subsection{Fully reconstructed jets}

In vacuum, jets are defined by algorithms combining hadrons or calorimeter towers in a certain way. The algorithms are deisgned to undo the pQCD evolution of a parton shower, such that jets on the detector level (calorimeter towers) are approximately equal to jets on the hadron level (particle tracks) and jets on the parton level (the output of pQCD calculations). This works because in the absence of a medium the flow of energy and momentum must always remain inside the jet, and allows to study the pQCD dynamics on various levels, from global event characteristics such as thrust to differential momentum distributions of hadrons inside jets.

However, this assumption is no longer true in the presence of a medium, as the perturbatively evolving parton shower in general exchanges energy and momentum with the non-perturbative medium. As a result, there is no unambiguous notion what a jet in medium should be: If one defines the perturbative part of the shower as jet, then the energy of jets will be smaller than the energy of the original parton, as part of its energy may be carried by non-perturbative excitations of the medium. If a jet is defined as everything causally correlated with the shower-initiating parton, then the jet energy will be in general higher than the parton energy, as collisions with medium partons may correlate them with the jet, while part of their energy is thermal. Finally, if the jet is taken to be the flow of energy and momentum of the original parton, then the jet energy will be the parton energy, but a jet can no longer be defined at the hadron level. 

The application of jet finding algorithms in a medium faces several complications \cite{JetMedium} which are related to the need to account for fluctuations in the background underlying the jet. Once jet properties are modified as compared to the vacuum case, additional complications appear. For instance, in \cite{YaJEM-Jets} is has been shown that imposing $P_T$ cuts (or to a lesser degree angular cuts) may significantly obscure the identification of the original parton energy scale.

The most striking manifestation of jet modification in a medium is the appearance of monojets which has been measured by the ATLAS and CMS collaborations \cite{ATLAS,CMS} and quantified in terms of the dijet asymmetry. The dijet asymmetry is a very complicated observable, as it requires to account for jet finding and jet energy identification in the presence of a fluctuating background medium as well as various biases introduced by the presence of a trigger jet in the event.

\section{Assumptions underlying Jet Quenching Models}

\label{S-Assumptions}

Usually it is tacidly assumed in the literature that whenever a calculation of a particular jet quenching model is shown together with experimental data this would test the parton-medium interaction model. However, this is not the case.

For the purpose of this discussion, let us define a parton-medium interaction model as a way to calculate the medium-modified fragmentation function (MMFF) $D_{i \rightarrow h} (z, E, Q_0^2 | T_1(\zeta), T_2(\zeta), \dots T_n(\zeta))$, i.e. the distribution of hadrons $h$ given a parton $i$ with initial energy $E$ and initial virtuality $Q_0^2$ where the hadron energy $E_h = z E$ and the parton has traversed a medium along the path $\zeta$ where $T_i(\zeta)$ are the medium transport coefficients relevant for the process. 

In this formulation, the tomographic information about the medium spacetime evolution is now in the $\zeta$ dependence of the $T_i$, whereas the information about the nature of parton-medium interactions is in the $T_i$ dependence of $D_{i \rightarrow h}$, where the actual evaluation involves suitable integrals of $T_i(\zeta)$ along the path $\zeta$.

This essential core of the parton medium interaction model is not experimentally testable, since we can neither prepare initial partons with given  $(E,Q^2)$, nor can we prepare a medium with known and controlled $T_i(\zeta)$. Instead, parton type, initial parton kinematics, medium spacetime geometry and initial parton position inside this geometry are event by event random variables of which we at best know the probability distribution. Any comparison of a parton-medium interaction model thus requires to embed this model into a larger framework which supplies these probability distributions.

Thus, when we talk about for instance 'testing a radiative energy loss scenario' there are many assumptions made both inside the particular framework of modelling radiative energy loss and in the surrounding framework.

\subsection{Assumptions in the general framework}

Generically, the framework in which a parton-medium interaction model is tested can be characterized as follows: The initial hard process is computed using pQCD. Apart from the transition from nucleon parton distribution functions (PDFs) (e.g. \cite{CTEQ1,CTEQ2}) to nuclear PDFs (e.g. \cite{NPDF,EKS98,EPS09}) to take into account changes in the partonic structure of the initial state, the hard process itself is taken to be identical in vacuum and medium. This can be justified by the uncertainty relation comparing the inverse hard scale with the inverse temperature of the medium to estimate the relevant formation times and is at the heart of the idea of tomography: If one could not compute the initial parton spectrum before passage through the medium reliably, one could not make any conclusions with regard to the strength of the final state interaction with the medium from the measured spectra in p-p and A-A collisions and would not have tomographic information.

The hard partons from the primary process are now embedded into a medium at an initial vertex position $(x_0,y_0)$ in the plane transverse to the beam axis with a random orientation $\phi$ with respect to the reaction plane of the A-A collision. The medium description needs to specify  $T_i(\zeta)$ for $\zeta \sim (x_0 + \tau \cos(\phi), y_0 + \tau \sin(\phi))$ where $\tau$ is the evolution proper time. However, in most cases the medium evolution model is taken to be a relativistic fluid dynamics code with parameters tuned to describe bulk medium properties (e.g. \cite{hydro2d,hydro3d}). In this case, a model $T_i(T, \epsilon, s,\dots)$ connecting the termodynamical variables like temperature $T$, energy density $\epsilon$ or entropy density $s$ which appear in the hydrodynamical framework with the transport coefficients $T_i$ as needed to obtain the MMFF must be specified.

After propagating the partons through the evolving medium, the MMFF can be obtained and convoluted with the primary parton spectrum to yield the final hadron spectrum. This procedure needs to be averaged over all possible initial vertices and parton orientations. After the final hadron distribution is corrected for experimental cuts, comparison with data can be made. Most observables are ratios of in-medium over in-vacuum quantities, these also require a baseline computation in which the MMFF is replaced by the vacuum fragmentation function obtained in the same parton-medium interaction model for the limit of vanishing medium.

Let us now review a number of model assumptions in this chain which are not related to specifics of the parton-medium interaction and what is known about their validity.

\subsubsection{Applicability of LO pQCD for the hard process}

In most models, the initial hard process is computed in LO pQCD (supplemented with a phenomenological $K$-factor to account effectively for higher order processes) with the underlying assumption that the accuracy of such a calculation is sufficient and that the error caused by neglecting higher order processes is small as compared to other uncertainties in the calculation.

Since observables such as $R_{AA}$ or $I_{AA}$ are ratios of spectra where any $K$-factor in the calculation drops out, the relevant question is if a LO pQCD calculation can describe the shape of hard hadron spectra with sufficient accuracy to serve as baseline. If this is the case, then one can argue that the hard process itself is described with sufficient accuracy and all medium modifications can be safely absorbed into the MMFF. In \cite{LO-pQCD} it has indeed been shown that a reasonably good description of high $P_T$ hadron spectra can be done with LO-PQCD and a $K$-factor.

Note that this question is similar to the validity of PYTHIA \cite{PYTHIA} for the description of high $P_T$ hadron spectra. PHYTIA is a LO pQCD framework supplemented with the PYSHOW  algorithm \cite{PYSHOW} to model a parton shower and the Lund model for hadronization. The latter two in combination are nothing but a model to compute the fragmentation function and its scale dependence. It is known that the simulation of a parton shower captures many features of NLO pQCD, but fails in certain kinematical situations such as truly hard gluon radiation (3-jet events). A similar argument can be made here: An LO pQCD baseline with a sufficiently well-computed MMFF should be valid to good accuracy except for specific event topologies where NLO pQCD is explicitly important.

Energy loss for a primary process computed in next-to-leading order pQCD has been studied e.g. in \cite{NLO-Dihadrons,NLO-gamma-h}, however with rather drastic assumptions (see below for a discussion) about the medium geometry (hard sphere overlap, no transverse expansion) and parton medium interaction (mean energy loss only, no fluctuations) that the added value of the NLO baseline is difficult to assess. 

\subsubsection{The spacetime distribution of the medium}

The spacetime distribution of matter produced in heavy-ion collisions is ideally something one would like to extract in a model-independent way from the experimental measurements. However, it is now known that this is not possible due to the many averaging steps need to compute observable quantities --- even simpler inversion problems turn out to be ill-defined \cite{Inversion}.

Thus, any comparison of a model with data requires to specify a suitable guess for $T_i(\zeta)$ \emph{a priori} and constitutes a test of (among other things) how well a particular combination of parton-medium interaction model and medium model describes the data. If one could show that the outcome of the computation is largely independent of the medium modelling, then one would not have any tomographic sensitivity but could test the physics of parton-medium interaction well. If, on the other hand, the results of all reasonable parton-medium interaction models would be largely generic given a specific medium model, then one could do tomography. In practice, neither is the case. 

In \cite{RAA-hyd3d} it was demonstrated that one could obtain an equally good fit to the $P_T$ dependence of $R_{AA}$ in 200 AGeV central Au-Au collisions using the same parton-medium interaction models with three medium evolution models constrained by bulk data (i.e. incorporating longitudinal and transverse matter expansion such that bulk low $P_T$ hadron spectra were reproduced correctly). However, the extracted medium parameters were a factor two different between models. Thus, if one would have started with a given model for $T_i(\epsilon)$ and simply used the energy density as given by the medium model, the three scenarios would have resulted in vastly different suppression of high $P_T$ hadrons. In the same work, the case of an (unphysical) static medium was also analyzed --- here the medium parameter was a factor 5 different from the evolving case. This establishes that simplistic medium models (such as e.g. the uniform cylinder decaying after a finite time as used in \cite{Fragility}) can at best yield an order of magnitude estimate of the medium transport coefficients.

The tomographic sensitivity of high $P_T$ hadrons to properties of the evolving medium is further illustrated in \cite{RAA-hydro-systematics} at the example of the spread $S^{in}_{out}$ between in-plane and out of plane emission of high $P_T$ hadrons. It was found that the spread shows a sensitivity to the choice of the medium evolution model which is of the same order of magnitude than the sensitivity to the choice of the parton-medium interaction model. This means that the success or failure of a particular parton-medium interaction model to describe data in many cases can not be discussed independently from the chosen medium model. In general, only combinations of models can be tested against the data. Thus, every meaningful test of a parton-medium interaction model \emph{must} involve a systematic study with different well-constrained medium evolution models.

\subsubsection{Relevance of hydrodynamical timescales}

Hydrodynamical codes (e.g. \cite{hydro2d,hydro3d,hydro-Hirano,vhydro-Song-Heinz}) are a tool to evolve a thermalized initial state to a final state. The thermalized initial state is usually taken to be present at a proper time $\tau_i$, the equilibration timescale of the system. The endpoint of the hydrodynamical evolution is chosen to be an isosurface, often an isothermal at $T_F$, the so-called freeze-out temperature, at which point the Cooper-Frye prescription \cite{Cooper-Frye} is used to convert the hydrodynamical fluid into a distribution of hadrons.

When embedding hard partons into the background of an evolving medium, it is often assumed that $\tau_i$ and $T_F$  are the relevant parameters for the jet as well, but there is no compelling reason why this should be the case. Hard partons do not probe the degree of isotropization in the soft medium or pressure gradients, rather they probe the density of coloured scattering centers at a certain resolution scale. Thus, a hard parton my well scatter from a pre-equilibrium state, or interact with a hadron which is free-streaming from a hydro-perspective.

This problem was investigated in \cite{RAA-hydro-systematics} with the result that for the most common class of energy loss models (radiative energy loss with a decoherence condition, see section \ref{S-PartonMedium}), the uncertainties are parametrically small. In practice, they were found to be O(15\%) in the extraction of medium parameters and the spread in $R_{AA}(\phi)$, i.e. smaller than the variation between different hydrodynamical models O(100\%). 

More advanced hydrodynamical codes do no longer use the Cooper-Frye prescription to compute the final hadron spectrum, but rather run a hadronic cascade following the hydrodynamical decoupling (see e.g. \cite{hydro3d,VISHNU}). Such models are superior in describing bulk data, but to date there is no consistent scheme how to embed hard partons into the hadronic rescattering stage.

\subsubsection{Transport coefficients and thermodynamics}

When embedding a parton-medium interaction model into a hydrodynamical medium, a model for the relevant transport coefficients based on the values of hydrodynamical parameters needs to be specified. The effect of this model choice on the extraction of transport coefficients has been studied in \cite{RAA-eloss-systematics}.

For radiative energy loss, the relevant transport coefficient is $\hat{q}$ (the virtuality transfer from medium to parton per unit pathlength) which is commonly modelled as \cite{Flow1,Flow2}

\begin{equation}
\label{E-qhat}
  \hat{q}(\zeta) = K \cdot Q(\zeta) 
                 \bigl(\cosh \rho(\zeta) - \sinh \rho(\zeta) \cos\alpha(\zeta)\bigr)
\end{equation}

where $Q(\zeta)$ is the local density of scattering centers, $\rho (\zeta)$ is the local transverse flow rapidity of the medium (in the commonly assumed Bjorken-type longitudinal flow fields, a hard parton is always longitudinally co-moving with the medium at any given rapidity slice and never sees a flow difference with the medium) and $\alpha$ is the angle between parton trajectory and flow direction. $K$ is an overall free parameter to adjust the strength of the parton-medium interaction given a certain density of scattering centers.

Common choices for $Q$ are $\epsilon^{3/4}, T^3$ or $s$. For an ideal gas, all these prescriptions are identical and count the entropy content of the medium, whereas the second part of the equation corrects for the Lorentz contraction of the volume induced by the relativistic flow, and hence the apparently higher density of scatterers as seen by the propagating parton. However, in the presence of a phase transition and a hadronic evolution phase, the three prescriptions lead to sizable differences in the late stage. $\hat{q} \sim T^3$ significantly emphasizes the hadronic phase as compared to $\hat{q} \sim \epsilon^{3/4}$ \cite{RAA-eloss-systematics}. In the extraction of $\hat{q}$, this leads to a factor 2 difference.

Based on these results, it appears that the question how the parton-medium interaction is treated in a hadronic medium is of some importance at the level of quantitative parameter extraction. 

\subsubsection{Smoothness of the medium}

In recent times, it has become apparent that event by event (EbyE) fluctuations in the hydrodynamical initial state are an important effect. In other words, it matters for observable quantities whether one averages first over many initial states and computes the final state by hydrodynamically evolving an average, smooth state, or if one evolves many fluctuating irregular initial geometries and averages observables over the resulting final states. This has driven the development of EbyE hydrodynamical codes \cite{VISHNU,fhydro-Holopainen,fhydro-Schenke,fhydro-Heinz}

This development naturally raises the question to what degree the modelling of partons propagating through a medium is influenced by choosing a final state rather than an initial state averaging procedure. First studies in \cite{feloss-Fries1,feloss-Fries2} using a non-evolving geometry identified two main effects, the non-linear response of $R_{AA}$ to 'holes' and 'hotspots' in the density distribution and the correlation of initial hard parton production vertices with hotspots with large differences of factors $\sim 2$ in the extracted transport coefficient between initial state averaged, smooth medium and final-state averaged EbyE fluctuating medium. A subsequent calculation using an ideal fluctuating hydrodynamics code \cite{fhydro-Holopainen} showed that the two effects almost cancel in practice and that the net difference in the transport coefficient between smooth and averaged medium is rather O(20\%) \cite{feloss-Renk}. A large reason for this difference to \cite{feloss-Fries1,feloss-Fries2} is that the medium evolution itself removes initial state fluctuations quickly, as they cause large pressure gradients. This is even more true in a viscous evolution of initial state fluctuations (see e.g. \cite{fhydro-Schenke}).

There is, however, a residual effect of fluctuations for the normalization of $R_{AA}$ for non-central collisions which depends on the size scale of the fluctuations \cite{feloss-Renk}. This requires further study with more systematics.

\subsubsection{Spatial distribution of hard vertices}

The probability distribution for a hard vertex to be found in the transverse plane at some position $\bm{r}_0= (x_0,y_0)$ is usually taken to be the binary collision profile for impact parameter $\bm{b}$

\begin{equation}
\label{E-Profile}
  P(x_0,y_0) = \frac{T_{A}(\bm{r}_0{+}\bm{b}/2) T_A(\bm{r}_0{-}\bm{b}/2)}
                    {T_{AA}(\bm{b})},
\end{equation}
where  the nuclear thickness function is given in terms of the Woods-Saxon 
nuclear density $\rho_{A}(\bm{r},z)$ as 
\begin{equation}
  T_{A}(\bm{r}) = \int dz\, \rho_{A}(\bm{r},z).
\end{equation}

The relevance of this choice becomes apparent only in connection with the hydrodynamical initial state. Consider the case of a cylinder of constant density with the Woods-Saxon radius $R$ as medium model. Embedding a binary overlap profile in such a distribution creates through the tail of the binary profile a halo of partons  which never encounter the medium at all. Similarly, when a more realistic wounded nucleon (WN), binary collision (BC) or color-glass-condensate (CGC) profile is chosen for the initial state, this choice influences the size of the halo of non-interacting partons \cite{RAA-hydro-systematics}.

As mentioned above, in the case of fluctuating initial conditions, it is particularly important that the same set of vertices is used in an event to both generate the hydrodynamical initial state and to sample the hard parton production points, otherwise the real correlation between hotspots in the initial density and initial hard parton position is lost \cite{feloss-Fries1,feloss-Fries2,feloss-Renk}.

\subsubsection{Eikonal parton propagation}

In many implementations of parton-medium interaction, it is assumed that partons probe the medium on straight line trajectories. However, interactions with the medium in general transfer energy and momentum and hence deflect the trajectory of the parton. Generically, one would always expect the hard scale to be larger than the temperature, $p_T \gg T$, and hence the actual amount of deflection should be small for sufficiently hard partons.

In \cite{elasticMC1} the eikonal propagation assumption was tested explicitly and found to be good on the 10\% level as far as the origin of observed partons in the final state is concerned. It should be noted however that non-eikonal deflection of hard partons has no effect on single hadron high $P_T$ observables unless the deflection is strong enough to probe appreciably different regions in a hydrodynamics or parton kinematics. For instance, in a Bjorken picture, the loss of partons produced at midrapidity and being scattered to forward rapidity would be compensated by a corresponding gain from partons produced at forward rapidity being scattered into midrapidity.

Non-eikonal parton propagation is experimentally accessible in dihadron correlations.

\subsubsection{Summary}

The following table summarizes the uncertainties associated with various assumptions as estimated above (percentages are always given with reference to the lowest value observed in a systematic study). We distinguish two cases: 1) the extraction of a transport coefficient $T_i$ like $\hat{q}$ or $\hat{e}$ (i.e. a measure of the overall strength of parton-medium interaction) and 2) a less averaged observable such as the spread $S^{in}_{out}$ between in-plane and out of plane emission at high $P_T$, which is the ratio of suppression factors in which many systematic uncertainties cancel.

\begin{table}[htb]
\begin{tabular}{|l|r|r|c|}
\hline
& $\hat{q},\hat{e}$ & $S^{in}_{out}$ & references\\
\hline
LO pQCD & 10\% &  $<$5\%  & \cite{LO-pQCD}\\
medium model & 100\% & 100\% & \cite{RAA-hyd3d,RAA-hydro-systematics}\\
hydro timescales & 15\% & 15\% & \cite{RAA-hydro-systematics}\\
$T_i(T,\epsilon,s)$ model & 100\% & $<$5\% & \cite{RAA-eloss-systematics}\\
medium smoothness & 20\% & $<$5\% & \cite{feloss-Renk}\\
eikonal propagation & $<$5\% & $<$5\% & \cite{elasticMC1}\\
\hline
\end{tabular}
\end{table}

Two observations can be made immediately: First, there is a hierarchy in the sensitivity: Not all model choices affect the results in the same way. To a first approximation, we may thus focus on the quantities to which the results are most sensitive. Second, by forming suitable ratios of more differential observables, the sensitivity can be focused to one assumption class which can then potentially be determined from this particular observation.

How to interpret a 100\% sensitivity of $\hat{q}$ or the spread to the choice medium model is in the eye of the beholder. If one is interested in extracting transport coefficients or the physics of energy loss in a model-independent way from data, then these numbers imply that this can not be done in a meaningful way. On the other hand, a large sensitivity to the hydrodynamical bulk medium model means that there is tomographic sensitivity in the observable and hence a chance to constrain different bulk models through high $P_T$ physics.

Let us stress again that these estimates assume a \emph{constrained} model for the medium including a realistic time evolution and that the sensitivity to model parameters is easily an order of magnitude larger if unconstrained, static or otherwise unrealistic medium models are used. This may explain why unconstrained models allow for spectacular effect sizes which decrease dramatically once the calculation is repeated in a constrained framework.

\subsection{Assumptions about parton-medium interactions}

\label{S-PartonMedium}

Let us continue with a discussion of assumptions needed to create a model for parton-medium interaction.

\subsubsection{Basic interaction structure}

\label{SS-Interaction}

At the core of any parton-medium interaction model is an idea by what processes energy is carried away from the hard parton, and commonly the models are labelled according to this idea. 

The simplest possibility are elastic QCD interactions with medium constituents, in which the recoil energy of struck thermal partons is lost from the hard parton \cite{elastic1,elastic2,elastic3,elastic4,elastic5,elastic6}. Such models have been suggested as an attempt to understand the surprisingly large suppression of heavy quarks \cite{HQPuzzle}, but the arguments apply equally well to light quarks and gluons. Key quantity determining the magnitude is the effective mass of the recoiling objects --- under the assumption that the medium consists of static scattering centers, there is no collisional energy transfer. Modern formulations of the same idea treat the problem in MC codes \cite{elasticMC1} or by solving coupled Fokker-Planck equations numerically \cite{elastic-Schenke}.

The distinctive feature about elastic interaction is that they are \emph{incoherent}. If a parton traverses a medium with constant density $\rho$ and looses the mean energy $\langle \Delta E \rangle_{1 scatt}$ in each scattering processes, then the mean energy loss after a length $L$ will be 
\begin{equation}
\langle \Delta E \rangle = N_{scatt} \langle \Delta E \rangle_{1 scatt} = \sigma \rho L \langle \Delta E \rangle_{1 scatt} \sim L
\end{equation}

(with $\sigma$ being the interaction cross section) as long as finite energy corrections are not important, i.e. as long as $\langle \Delta E \rangle_{1 scatt}$ can be taken independent of the initial hard parton energy $E$.

The mean energy loss per unit pathlength is often cast into the form of the transport coefficient $\hat{e}$ and its variance into $\hat{e_2}$ \cite{JetReview}.

A different possibility to transport energy away from a hard parton is medium-induced radiation \cite{radiative1,radiative2,radiative3,radiative4,radiative5,radiative6}. The underlying idea is that as the hard parton traverses a medium, interactions change the transverse momentum randomly. The transport coefficient $\hat{q}$ measures the rate $dQ^2/dx$ at which a parton acquires virtuality from the medium per unit pathlength by this mechanism. A gluon from the virtual cloud surrounding the hard parton can decohere from the parent and appear as real radiation if the interaction with the medium is sufficient to overcome the cloud gluon's virtuality \cite{QuenchingWeights}.

Assuming a parent with energy $E$ and a radiated gluon with energy $E_{rad}$ and momentum $k_T$ transverse to the parent where $E \gg E_{rad} \gg k_T$ (soft and collinear emission), the formation time associated with the radiation process as given by the uncertainty relation is $\tau \sim E/k_T^2$ or, assuming the transverse momentum is acquired by interactions with the medium, $\tau \sim E/Q^2$. Since this virtuality is picked up as $Q^2 = \hat{q} \tau$, the two expressions can be combined to find $\tau \sim E_{rad}/(\hat{q} \tau)$ which can be solved for the typical radiated energy $E_{rad}$. The process is most efficient when as much time for decoherence is available as possible, for a medium length $L$ and a parton moving with the speed of light this implies $\tau = L$, thus

\begin{equation}
E_{rad} \sim \hat{q} L^2 \sim T^3 L^2
\end{equation}

since $\hat{q}$ in a thermal medium is parametrically $\sim T^3$. This different power in the pathlength is often used to argue that radiative energy loss dominates over collisional energy loss.

Common model frameworks for this type of parton-medium interaction are the Armesto-Salgado-Wiedemann (ASW) formalism \cite{QuenchingWeights,ASW-1}, the Arnold-Moore-Yaffe (AMY) formalism \cite{AMY-1,AMY-2}, the Guylassy-Levai-Vitev (GLV) formalism \cite{radiative5,WHDG} and the higher twist (HT) formalism \cite{radiative6,HT-DGLAP}.

While early radiative energy loss model used the approximation that the parent parton has such large energy that the change in its kinematics induced by the radiation of a soft gluon can be neglected, finite energy and length effects for medium-induced radiation have now been worked out in a number of models including MC frameworks where explicit energy-momentum conservation at each vertex can be accounted for \cite{Korinna-LPM,YaJEM-D,Caron-Huot}. The conclusions of all these works agree qualitatively: While $L^2$ dependence of energy loss is observed initially, after a short timescale (typically 2-3 fm) kinematical and finite length correction effectively lead the system back to a linear dependence on pathlength. Thus, for realistic kinematics, the $L^2$ dependence of radiative energy loss does not lead to dramatic differences as compared with an incoherent mechanism.

There is yet a different class of scenarios based on the idea that the AdS/CFT correspondence \cite{AdS1,AdS2,AdS3} can be used to describe a QCD medium in the strongly coupled limit which yields a parametrically different estimate for the energy loss of a hard parton (if everything is treated in the strong coupling limit, no jets appear in the model). In so-called hybrid models, it is assumed that what is radiated into the medium still comes from the perturbative part of the hard parton wave function, but  that the interaction of the virtual gluons with the medium, and the way they are freed, is governed by strong-coupling dynamics \cite{hybrid1,hybrid2,hybrid3}.

Here, what acts on the gluon cloud is a longitudinal drag force of order $F \sim T^2$, thus the virtuality scale of gluons from the hard parton which is accessible after a propagation length $L$ is $Q^2 \sim (F \cdot L)^2 \sim T^4 L^2$. If this estimate is inserted into the expression for the formation time above, parametrically

\begin{equation}
E_{rad} \sim T^4 L^3
\end{equation}

is found. Thus, a strong coupling treatment of the interaction of the medium with the virtual gluon cloud of a hard parton is has both a different temperature and length dependence than a weak coupling treatment.

A very relevant question is how finite energy corrections would change the pathlength dependence found in this class of models. While there does not seem to be a detailed framework in which this question has been answered, comparison with the pQCD induced radiation models suggests that the pathlength dependence may be weakened after a finite time by at least one power of $L$. 

Finally, note that realistic models usually combine different parton-medium interaction scenarios \cite{WHDG,el_rad_amy,el_rad_yajem}, thus the real question is about the relavtive strength rather than about which mechanism is true.

\subsubsection{Medium-modified showers vs. energy loss}

\label{SS-Fragmentation}

As stated above, the general outcome of a parton-medium interaction model is the MMFF $D_{i \rightarrow h} (z, E, Q_0^2 | T_1(\zeta), T_2(\zeta), \dots T_n(\zeta))$ which describes the the QCD evolution to a final state in the presence of a medium after a highly virtual parton has been created in a hard process. In principle, this evolution includes the development of a partonic shower as well as the hadronization. In practice, existing in-medium shower codes like JEWEL \cite{JEWEL}, YaJEM \cite{YaJEM1,YaJEM2} or Q-PYTHIA \cite{Q-PYTHIA} use the formation time for hadrons $h$ with energy $E_h$  and mass $m_h$ to argue that $\tau \sim E_h/m_h^2$ corresponds, for light hadrons or sufficiently energetic hadrons, to a scale outside the medium geometry and hence employ vacuum hadronization models. It should be noted that this approximation is frequently not justified for baryon production or subleading shower fragments.

However, the majority of parton-medium interaction model makes the further approximation that it is sufficient to consider the energy shift of the leading parton, i.e. the MMFF is approximated by

\begin{equation}
D_{i \rightarrow h} (z, E, Q_0^2 | T_i(\zeta)) \approx P(\Delta E,E|T_i(\zeta)) \otimes D_{i \rightarrow h}(z, Q_0^2)
\end{equation}

where $D_{i \rightarrow h}(z, Q_0^2)$ is the vacuum fragmentation function and $P(\Delta E,E)$ is the probability distribution for the leading parton to experience an energy shift $\Delta E$ before fragmentation.

\emph{A priori} it doesn't seem obvious that this approximation works, since the high virtuality scale $Q_0$ after the hard process ensures very fast partonic branching even before a medium can be formed, thus the object interacting with the medium is almost never a single on-shell parton but rather a partially developed shower, and it is far from obvious that the vacuum QCD evolution can be factored out and carried out after the medium-induced energy loss has been considered. 

However, if one convolutes a vacuum fragmentation function with a pQCD parton spectrum at RHIC kinematics, one finds a mean $z$ of $\sim 0.7$, i.e. for the \emph{observed} single inclusive hadron spectrum, about 70\% of the original energy are in the leading hadron. One can combine this with a Lund-type hadronization model \cite{Lund} to estimate that this implies that on the partonic side more than 70\% of the shower energy flow through the leading parton, thus the \emph{single inclusive} hard hadron spectrum is dominated by events in which there is little vacuum radiation and the hadron kinematics is driven by the leading parton kinematics. In this particular case, it is to good accuracy justified to approximate the situation by leading hadron energy loss. Note that this argument does not apply to different observables such as dihadron correlations or jets where the energy loss approximation is bound to fail for subleading jet fragments.

From another perspective, this corresponds to different ways of interating the basic parton-medium interaction process. In in-medium shower evolution codes \cite{JEWEL,YaJEM1,YaJEM2,Q-PYTHIA} the parton-medium interaction is embedded via a spacetime picture of the evolving shower into the iteration of basic $1\rightarrow 2$ splitting processes with explicit energy/momentum conservation (inclusing the transverse momentum) at each vertex. A similar setup is used in the more recent version of the HT formalism \cite{HT-DGLAP} where the basic parton-medium interaction is part of the (longitudinal) DGLAP evolution of the fragmentation function. As we will see later, there are reasons to think that the model choice here mainly affects the distribution of subleading hadrons in the fragmentation function. 

The AMY formalism \cite{AMY-1,AMY-2} starts with the assumption of a hard on-shell parton, but evolves the whole parton distribution created by medium-induced radiation with rate equations. Finally, in older formulations of HT \cite{radiative6} or the ASW formalism \cite{QuenchingWeights} the assumption of independent gluon emissions is made which allows to convolute the gluon spectrum of a single parton-medium interaction with the Poisson distribution to compute the total energy loss probability density.

There are also hybrid models such as MARTINI \cite{MARTINI} in which a vacuum shower is evolved till a medium is formed, and the energy loss approxiation is then used for each of the shower partons present at this point.

In formulations in which the parton-medium interaction is formulated as part of an evolving shower from some initial hard scale $Q_0$ down to a lower scale $Q_F$, there is an additional source of pathlength dependence. It can be argued that the uncertainty relation permits the medium of length $L$ to modify the shower resulting from a parton with energy $E$ only down to a scale $Q_F$ given by \cite{JetReview,HT-DGLAP}

\begin{equation}
\label{E-Q0}
Q_F = \sqrt{E/L}.
\end{equation}

This condition amounts to an additional energy and length dependence of the MMFF which can't be easily cast into an analytical expression and has been used in the resummed HT framework \cite{HT-DGLAP} and in the MC code YaJEM-D \cite{YaJEM-D}.

\begin{figure}[htb]
\epsfig{file=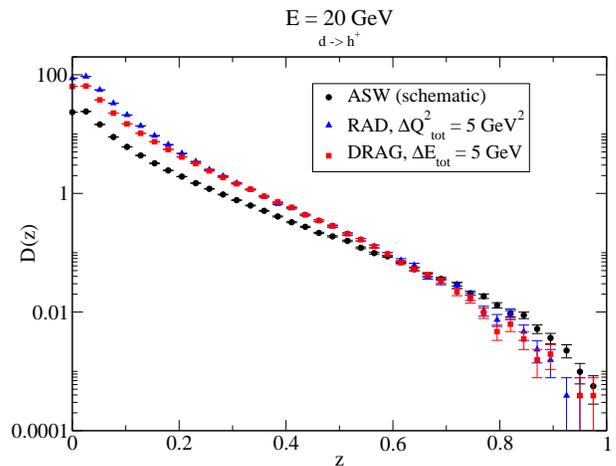, width=8cm}
\caption{\label{F-MMFF-eloss-shower} A comparison of medium-modified fragmentation functions of a 20 GeV $d$-quark as computed in the energy loss formulation (ASW) and an in-medium shower evolution (YaJEM with radiative energy loss (RAD) or incoherent drag (DRAG)) with the condition that all MMFF lead to the the same single hadron suppression \cite{YaJEM2}.}
\end{figure}

As evident from Fig.~\ref{F-MMFF-eloss-shower}, the differences between a MMFF computed in a leading parton energy loss picture and a MMFF from an in-medium shower evolution can be quite substantial. Especially at low $z$, the in-medium shower treatment keeps explicitly track of hadron production from he medium-induced soft gluon radiation whereas the leading parton energy loss does not. Based on the figure, one may wonder why the energy loss approximation works at all. The reason is that the MMFF \emph{for single inclusive hadron spectra} needs to be convoluted with the pQCD parton $p_T$ spectrum, and this convolution suppresses the high and low $z$ region. It turns out that for RHIC kinematics, the MMFF in this example is probed dominantly around $z\sim 0.6$ where any differences are small. This suggests that the suppression of single inclusive hadron production in terms of $R_{AA}$ is particularly unsuited to probe details of the MMFF and that more differential correlation observables such as $I_{AA}$ need to be studied to access this information.

\subsubsection{Mean energy loss vs. fluctuations}

If one assumes that the whole parton spectrum is shifted by a mean energy loss $\langle \langle \Delta E \rangle \rangle$, one can describe the hadron suppression observed at RHIC with a value of about $\langle \langle \Delta E \rangle \rangle \sim$ 5 GeV \cite{gamma-h}. However, such an exercise is not particularly meaningful, as it extracts a quantity both averaged over geometry (i.e. the tomographical information) and dynamical fluctuations given a path in the geometry (i.e. the microscopical dynamics of the parton-medium interaction), which is in addition biased by the fact that a hadron has been observed at high $P_T$, whereas there is no guarantee that all hard partonic scatterings lead to observed hard hadrons.

In some computations, the geometrical averaging has been undone but the dynamical fluctuations are averaged over. This corresponds to a fixed value of energy loss given a path, but since partons travel on different paths through the medium, the energy loss acquires a probability distribution which is then convoluted with the pQCD parton spectrum \cite{NLO-Dihadrons,NLO-gamma-h}. A direct comparison \cite{QM09proc} of model results with a computation retaining the dynamical fluctuations suggests that the tomographical interpretation of the result is drastically different. While in an average picture parton propagation through the medium center is strongly suppressed, strong fluctuations make it a likely scenario.

If one considers the strength of the fluctuaions, there is no compelling reason to assume that dynamical fluctuations can be averaged out. Plots of $P(\Delta E)$ in radiative energy loss formalisms typically do now show a narrow peak around the mean energy loss, but a wide, non-Gaussian distribution probing the whole available range of energies \cite{Brick}. For incoherent processes, this has been also demonstrated in a realistic treatment of elastic energy loss \cite{elastic-Schenke} where it was also found that the dynamical fluctuations are strong and non-Gaussian. 

\subsubsection{Implementation details}

Different implementations of the same physics idea can differ in parameters such as cutoffs, regulators to treat kinematical situations outside the model assumptions and similar details. A very instructive review comparing the ASW, GLV, AMY and HT formulations of radiative energy loss in a fixed-length constant density medium ('brick') illustrates the effect of these details \cite{Brick}, and we refer the reader to this work for an in-depth discussion of these effects. To summarize the findings very briefly: In an extreme case, the same observed suppression pattern would lead to a factor 8.5 (!) difference in the extraction of the medium transport coefficient dependent on which formulation is used. This uncertainty is largely driven by implementation-specific details and hence is present in addition to all the uncertainties not related to the parton-medium interaction as discussed in the previous section.

\subsubsection{Summary}

The sensitivity to assumptions discussed in this section fall into rather different categories. Some are directly linked to research questions, for instance the mass of the medium constituents probed by a hard parton which regulates the relative strength of elastic energy loss is not just a model parameter, but a probe characterizing the mircoscopical degrees of freedom in the medium, i.e. a quantity to be determined by comparison with experiment. Qualitatively different dynamics based on e.g. the choice of the parton-medium interaction is best not discussed in terms of uncertainties but in terms of different {\em a priori} plausible scenarios to be compared with data.

The sensitivity to other assumptions should not be seen as uncertainty in the modelling --- if an approximation of constant energy loss given a path results in a qualitatively different picture of the process than a more detailed computation allowing for fluctuating energy loss around its mean value, then this demonstrates just that the approximation is not justified. Similarly, wherever a full in-medium shower evolution shows qualitative differences to an energy loss calculation, this signals the breakdown of the energy loss approximation rather than an uncertainty in the calculation.

The sensitivity to implementation specific details as demonstrated in \cite{Brick} is most troublesome, as this (in the absence of a clear picture of a hierarchy in the assumptions) indicates that the approximations may be fundamentally flawed and too simplistic. However, these uncertainties mainly seem to influence the overall strength of the parton-medium interaction. In  \cite{RAA-hyd3d} it is demonstrated for ASW, AMY and HT that despite the huge discrepancies in the extracted medium parameter, once the interaction strength is fixed for each model to the data in one place, the resulting dynamics in terms of $P_T$ and centrality dependence or even unobservable quantities like hadron escape probability are similar to the 10\% level.

This, taken together with the observations of the preceding section that the uncertainties associated with a particular model assumption affect various observables in a very different way, gives rise to the hope that by testing models against a specific set of observables which are dominantly sensitive to a generic feature of models, one can constrain the physics of jet quenching in a meaningful way without overly being affected by specific implementation details. In other words, an incoherent model of parton-medium interaction will show a linear pathlength dependence of energy loss under very general conditions, quite independent of implementation details. If it can be demonstrated that a particular observable probes pathlength dependence and rules out linear dependence, then a whole class of models can be ruled out without reference to specific details.

\section{Constraining parton-medium interaction physics}

\label{S-Constraints}

In the following, the aim is to discuss constraints from a set of observables, each of which is in particular sensitive to few or ideally one generic model property. Thus, each of the observables provides different qualitatively different constraints which can be used to rule out classes of model with certain properties. In the following we will argue that the data on $R_{AA}(P_T, \phi, \sqrt{s})$ and $I_{AA}(p_T, \sqrt{s})$ constitutes such a set of observables.

Among these observables,  $R_{AA}(P_T)$ at RHIC kinematics exhibits a fairly generic $P_T$ dependence, but can be used as a starting point to fix all model parameters given the choice of a particular hydrodynamical model. The dependence of $R_{AA}$ on the reaction plane angle $\phi$, in particular the spread $S^{in}_{out}$ between emission into the reaction plane and out of the reaction plane is then, to almost equal parts, sensitive to the choice of the hydrodynamical model and to the pathlength dependence of the parton-medium interaction. The requirement that the same model describes $R_{AA}(P_T)$ at LHC energies of $2.76$ ATeV when a carefully constrained extrapolation of the medium model to larger $\sqrt{s}$ is done then probes the parton-medium interaction model-specific probability to find an almost unmodified fragmentation pattern. Finally, $I_{AA}$ can be used to observe the more differential pattern of energy redistribution for subleading shower hadrons once the pathlength dependence is fixed.  

It needs to be stressed again that the large medium evolution model dependence of observables makes it mandatory to use a medium evolution model which is constrained by bulk data.

\subsection{$P_T$ dependence of $R_{AA}$ at RHIC}

\label{SS-PTRHIC}

The $P_T$ dependence of $\pi^0$ $R_{AA}$ in central 200 AGeV AuAu collisions as obtained by the PHENIX collaboration \cite{PHENIX_RAA_phi} is usually the first high $P_T$ observable to be discussed in models.

For a wide range of model assumptions, the $P_T$ dependence of $R_{AA}$ in this kinematical situation is generic: In \cite{RAA-eloss-systematics} it is shown that no large difference between the ASW, HT and AMY formulation of energy loss can be seen and in \cite{RAA-hyd3d,RAA-hydro-systematics} no visible dependence on the assumptions about medium modelling are observed. An explanation for this finding is provided in \cite{gamma-h} by an analysis how  sensitive the convolution of parton spectrum and energy loss probability density $P(\Delta E)$ is to the choice of $P(\Delta E)$.

The main findings can be illustrated in a simple model. Consider suppression in the leading parton energy loss picture on the partonic level only.
$P(\Delta E)$ can then have a discrete transmission piece $T$ representing the probability for no energy loss and a continuous piece $P'(\Delta E)$ representing a parton shift in energy by $\Delta E$. After the convolution with a steeply falling parton spectrum, this is equivalent to a suppression since a parton which is after the energy shift found at $p_T$ must have originally come from a higher momentum $p_T + \Delta E$ where less partons are available. Here it is understood that $\Delta E > E$ implies that the parton is absorbed into the thermal medium.

$R_{AA}$ can then be understood from the ratio of modified over unmodified parton spectrum, where the modified parton spectrum at a given $p_T$ is determined by the number of partons escaping without energy loss plus the number of partons available in the spectrum at $p_T + \Delta E$ times the probability $P(\Delta E)$ of a shift by $\Delta E$. If we assume a power law $p_T^{-n}$ for the parton spectrum,
\begin{equation}
R_{AA}(p_T) \approx T + \int_0^{E_{\rm max}} d \,(\Delta E) P'(\Delta E)  \left(1+\frac{\Delta E}{p_T}\right)^{-n}.
\end{equation}
It is evident from the expression that $R_{AA}$ at a given $p_T$ is equal to the transmission term of zero energy loss plus a contribution which is proportional to the integral of $\langle P(\Delta E) \rangle_{T_{AA}}$ from zero up to the kinematic limit energy scale $E_{\rm max}$  {\it seen through the filter} of the steeply falling spectrum. $R_{AA}$ then generically grows with $p_T$ since the 'penalty factor' $\left(1+\frac{\Delta E}{p_T}\right)^{-n}$ for observing a parton downshifted by $\Delta E$ from a higher momentum scale decreases with increasing $p_T$ (in reality, the last statement is no longer true when the actual pQCD spectrum deviates significantly from a power law, i.e. around $\sqrt{s}/4$, however for realistic experimental conditions this region cannot be accessed with meaningful statistics).

 The speed of growth depends on the weight of $\langle P(\Delta E) \rangle_{T_{AA}}$ in the region from zero to $E_{\rm max}$ and on the power $n$ of the parton spectrum. 

At RHIC kinematics, the spectral power is $n \sim 7..8$ and in the limited kinematic range between 6 GeV where a perturbative description of the spectrum becomes available and 15 GeV where the experimental errors become substantial, a (moderate) 5 GeV shift in parton energy already corresponds to 80\% suppression. This means that the $P_T$ dependence of $R_{AA}$ in this domain only probes the probability that an energy loss model results in shifts between zero and $\sim 3..4$ GeV. In contrast, energy loss models typically expect shift probabilities throughout the whole allowed kinematic range $O(100)$ GeV \cite{Brick,Dihadron}. The implication is that the experiment is simply not sensitive to a large region of $P(\Delta E)$, instead fixing the normalization of models to data already limits the possible $P_T$ dependence to a range which is not distinguishable from a constant value given the experimental errors.

This means that any model which can be described effectively in terms of  $P(\Delta E)$ result in approximately the same $P_T$ dependence of $R_{AA}$, \emph{no matter} what model assumptions are made in detail. The obvious exception are models which have an explicit  dependence on the initial parton energy $E$ in the parton-medium interaction, i.e. can be effectively cast into the form $P(\Delta E | E)$. At least two such scenarios have been investigated: In \cite{gamma-h} it was established that a fractional energy loss $\Delta E = f E$ leads to a decrease of $R_{AA}$ with $P_T$ which is not supported by the data.  The resummed HT formalism \cite{HT-DGLAP} and YaJEM-D \cite{YaJEM-D} also contain, as discussed above, such an  explicit dependence which in this case leads to a stronger growth of $R_{AA}$ with $P_T$ than in other models.

It follows from the above that the only really non-trivial information from $R_{AA}(P_T)$ in central 200 AGeV collisions is the absolute normalization, i.e. the overall strength of the parton-medium interaction. Thus, given a chosen combination of medium evolution and parton medium interaction model, this data set can be used to e.g. fix the relation between thermodynamical parameters and transport coefficients, or to determine any other model-specific parameter which regulates the interaction strength. Once this is done, the combination of models to be tested is typically fixed for $\sqrt{s} = 200$ AGeV, i.e. different centralities at the same collision energy or different colliding ions can be computed without additional free parameters.

\subsection{Reaction plane angle $\phi$ dependence of $R_{AA}$ at RHIC}

\label{SS_ang}

The first non-trivial test of a parton-medium interaction model is then if it predicts the centrality dependence of $R_{AA}$ correctly. Note that hydrodynamical models are typically constrained using $v_2$ and bulk $P_T$ spectra across various centralities, thus while medium density and geometry changes from central to peripheral collisions, these changes happen within 'the same' hydrodynamical framework and are not arbitrary. In particular, they do not usually involve additional parameters which would affect the high $P_T$ side of the modelling.

Initially it was assumed that predicting the centrality dependence of $R_{AA}(P_T)$ averaged over reaction plane angle would be a reasonable test for models, and consequently this has been discussed in e.g. \cite{RAA-eloss-systematics}. However, note that the two endpoints are already constrained --- in central collisions, $R_{AA}$ is described by the models because this is the point where model parameters are adjusted (see above), $R_{AA}$ in ultra-peripheral collisions needs to go to unity because such collisions can not be different from a p-p collision, and models must monotonically connect these two limits, since both parameters relevant for the strength of energy loss, i.e. mean in-medium pathlength and mean density are monotonically decreasing from central to peripheral collisions. This means that studying angular averaged $R_{AA}$ is not a particularly sensitive test of models.

This is different if the angular dependence of $R_{AA}$ is taken into account and the suppression is studied as a function of the angle $\phi$ of a high $P_T$ hadron with the bulk matter reaction plane. For events inside a given centrality class, neither the mean density of the bulk medium nor the  mean geometry changes, but the mean pathlength which a particle spends in matter is very different between in-plane (i.e. along the short side of the almond-shaped overlap region) or out of plane (along the long side). Thus, the spread $S^{in}_{out}$ betwen in plane and out of plane $R_{AA}$ is a direct probe of how parton-medium interaction models respond to a change in in-medium pathlength.

Of course, $S^{in}_{out}$ does not only depend on the pathlength dependence of the parton-medium interaction, but also crucially on how large the difference in mean pathlength between in-plane and out of plane emission is, and this in turn is often linked to the diffuseness of the matter density profile --- for instance, in \cite{Diffuseness}, striking differences between a hard sphere and Gaussian overlap models were found.

Embedding a high $p_T$ event into a hydrodynamical bulk medium description never probes pathlength $L$ with the same power as in the constant medium case. The longitudinal Bjorken (e.g. \cite{hydro2d}) or close to Bjorken (e.g. \cite{hydro3d}) expansion implies that medium density drops like $\sim 1/\tau$ (where $\tau$ is the proper medium evolution time) and this effectively removes one power from the pathlength dependence. The transverse expansion velocity of the medium is initially small and grows over time as transverse pressure gradients lead to the development of flow. The resulting transverse expansion causes two potentially opposing effects --- as the medium expands, the distance to the relevant decoupling surface, and hence the in-medium pathlength may grow as compared to a transversally non-expanding case, whereas the medium density is decreased. In the case of a transversally expanding cylinder and $L^2$ dependence of energy loss, these two effects would exacly cancel. However, in any real system the dynamics depends on details --- the decoupling surface often moves inward in spite of transverse expansion, the rate at which density is diluted transversally depends on the stiffness of the equation of state, i.e. how efficiently spatial gradients are mapped into momentum space, and fluctuations in the initial state introduce further small-scale gradients and complicate the problem. Thus, there is no simple analytic way to understand how the mean pathlength difference between in-plane and out of plane is generated in a given hydrodynamical evolution.

For a representative set of four different hydrodynamical models, this problem has been studied in \cite{RAA-hydro-systematics}. Here, it was found that what matters most is late time dynamics in terms of how far out in terms of space and time the medium geometry extends, followed by initial dynamics, viscosity and initial profile diffuseness. The following table summarizes the findings:

\begin{table}[htb]
\begin{tabular}{|l|c|}
\hline
property of hydro & rel. sensitivity $S^{in}_{out}$\\
\hline
spatial size of freeze-out surface & $\sim$ 50\%\\
initial time scale & $\sim$ 25\%\\
viscous entropy generation & $\sim$ 18\%\\
initial profile diffuseness & $\sim$ 7\%\\
\hline
\end{tabular}
\end{table}

Thus, contrary to intuition, the actual choice of the initial matter density profile diffuseness is not a substantial effect. However, given this large variation of mean in-plane and out of plane pathlength between medium models which translates into a sensitivity of $S^{in}_{out}$ to the choice of a medium model which is as large as the sensitivity to the choice of a parton-medium interaction model \cite{RAA-hydro-systematics}, the implication is that pathlength dependence of energy loss can not be studied in a meaningful way unless the full systematics on the hydrodynamical side is accounted for.

As indicated above, the change of angular-averaged $R_{AA}$ as a function of centrality is constrained at the two endpoints of central and ultra-peripheral collisions. However, the evolution between these centralities depends both on the change in mean density and mean pahtlength, thus one would expect an ordering of models dependent on what power of the pathlength is probed by models, with the incoherend, $L$-dependent models responding slowest to a change in centrality and the AdS-type strong coupling $L^3$ dependent models responding relatively faster. Such trends are indeed observed when comparing parton-medium interaction models which each other \cite{RAA-hydro-systematics,elasticMCpathlength} however except in the case of linear pahtlength dependence they are not very significant with the data.

\begin{table*}
\begin{tabular}{l|ccccc}
model &  elastic $L$ & radiative  $L^2$ & AdS  $L^3$ & rad. fin. $E$ & min. $Q_0$\\
\hline
3+1d ideal & fails & works & fails &fails & works\\
2+1d ideal & fails & fails & marginal & fails & fails\\
2+1d vCGC & fails & marginal & works & fails & marginal\\
2+1d vGlb & fails & marginal & works & fails & marginal\\
\hline
\end{tabular}
\caption{\label{T-matrix}Viability of various combinations of medium model and parton-medium interaction model given the PHENIX data for $R_{AA}(\phi)$ \cite{PHENIX_RAA_phi} (based on calculations in \cite{RAA-hydro-systematics,YaJEM-D,elasticMCpathlength,elastic_phenomenology}, see text for the meaning of the various labels for models).}
\end{table*}

The combined results of \cite{RAA-hydro-systematics,YaJEM-D,elasticMCpathlength,elastic_phenomenology}, i.e. the viability of various physics mechanisms underlying a particular pathlength dependence given a choice of medium model and the data is best summarized in the form of a matrix (Tab.~\ref{T-matrix}).

It can be observed that two classes of models fail unconditionally, i.e. regardless what hydrodynamical model is chosen, i.e. the elastic (or incoherent) $L$-dependent type and the radiative type with finite energy corrections which for longer path is effectively $L$-dependent. The failure is quite dramatic --- $S^{in}_{out}$ differs from the data by at least a factor 6. For all other model types, it is possible to find a hydrodynamical evolution which results in a fair description of the data.

\begin{figure}[htb]
\epsfig{file=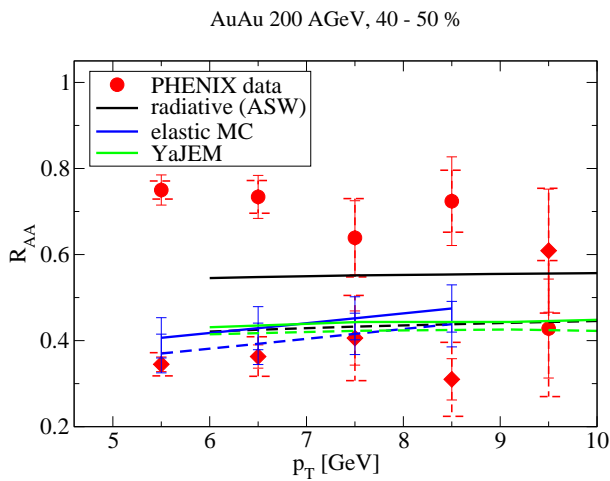, width=8cm}
\caption{\label{F-RAA-phi} Comparison of in-plane and out-of plane $R_{AA}$ for an elastic MC model  \cite{elasticMCpathlength}, YaEJM \cite{YaJEM-D} and the ASW radiative energy loss model \cite{QuenchingWeights} using the same 2+1d hydrodynamical medium evolution \cite{hydro2d} as compared with 40-50\% central 200AGeV AuAu collision data (circles: in plane, diamonds: out of plane) obtained by PHENIX \cite{PHENIX_RAA_phi}.}
\end{figure}

These findings are illustrated in Fig.~\ref{F-RAA-phi} at the example of the elastic energy loss MC calculation \cite{elasticMCpathlength}, YaJEM \cite{YaJEM1,YaJEM-D} and the ASW radiative energy loss model \cite{QuenchingWeights} embedded into a 2+1d ideal hydrodynamical model for the bulk medium evolution. With this particular medium model, radiative energy loss misses $S^{in}_{out}$ by about a factor two, but clearly reaches the right order of magnitude and hence is compatible with the data if the full uncertainty associated with the medium model choice is taken into account. In contrast, both the elastic model and YaJEM fail clearly in reproducing the spread and the average even allowing for the systematic uncertainty associated with the choice of the medium evolution.

The immediate conclusion is that a purely incoherent parton-medium interaction picture can be ruled out given the data. This is perhaps not surprising, as it has never been suggested that coherence is completely absent. However, given the fact that \emph{no model} overpredicts $S^{in}_{out}$ by more than 15\%, it can be inferred that an incoherent component must be  $\lesssim 10\%$ in order for a combined incoherent + other mechanism scenario to describe the data \cite{elastic_phenomenology,elasticMCpathlength}.

This is in contrast to straightforward computations of the relative magnitude of elastic energy loss from pQCD where frequently elastic components O(50\%) are found \cite{elasticMC1,WHDG,elastic_AMY}. The inevitable conclusion is that the main assumption made in pQCD calculations of elastic energy loss, i.e. that the medium DOF are almost free (quasi)-particles which can take a sizeable amount of recoil energy away from a leading parton does not appear to be true in nature.

It is also very instructive to note that YaJEM  in which the linear pathlength dependence arises only effectively due to finite energy and length corrections to a medium-induced radiation formalism which takes coherence into account and  which in principle exhibits $L^2$ dependence fails just in the same way as a formalism which is incoherent from the beginning \cite{YaJEM-D}. This indicates that the observable really probes pathlength dependence regardless of how it arises microscopically. The fact that finite energy corrections lead to the failure of a radiative energy loss formulation seriously questions the relative success of radiative energy loss models derived in infinite kinematics (such as ASW or the AdS model). However, this is a qualitative theoretical objection and as we shall see in the next section, these models can also be ruled out based on a comparison with data.

\subsection{$P_T$ dependence of $R_{AA}$ at the LHC}

Following the arguments given in \ref{SS-PTRHIC}, we can conclude that at LHC kinematics the harder primary parton spectrum (i.e. the lower power $n = 4..5$ in a power-law fit) will allow to observe a larger range of $P(\Delta E)$ and that thus the $P_T$ dependence of $R_{AA}$ in central 2.76 AGeV central PbPb collisions will probe, unlike in the RHIC case, model-specific differences. 

If one aims at a comparison of a model combination which is constrained by RHIC data with $R_{AA}$ at LHC, there are however a number of additional difficulties connected with extrapolating a hydrodynamical framework to higher $\sqrt{s}$ (for an extended discussion see \cite{RAA_LHC}). For instance, it is expected that for increasing $\sqrt{s}$ the particle production mechanism might change soft participant scaling for the produced entropy at low $\sqrt{s}$ to hard binary collision scaling at higher $\sqrt{s}$, thus even qualitatively changing the initial matter profile. Likewise, the matter equilibration time is expected to shorten, while at the same time the overall higher multiplicity would allow for a lower decoupling temperature for the bulk matter. Thus, unlike in the previous extrapolation to different centralities in the same colliding system, there is some reason to expect that additional free parameters may appear in the extrapolation of the hydrodynamical medium model from RHIC to LHC energies, some of which may be relevant for high $P_T$ physics as well. It is therefore unreasonable to expect that a RHIC constrained model would exactly match LHC data if the 'same' hydrodynamical model is used at both RHIC and LHC, since the definition of 'same' on the hydrodynamical side now hinges on details of the framework chosen to extrapolate the hydrodynamics. In the following, we employ a pQCD + saturation approach, the "EKRT model" \cite{EKRT} as the relevant framework to define the 'same' hydrodynamics for different $\sqrt{s}$ \cite{RAA_LHC}. 

Note that this specifically requires the use of the 2+1d hydrodynamical framework \cite{hydro2d} as the underlying medium model which works \emph{worst} with many parton-medium interaction models for angular differential observables at RHIC energies (see \ref{SS_ang} above). Thus, while we know that $R_{AA}(P_T)$ for central collisions is not very sensitive to the geometry of the hydrodynamical modelling, there is no reason to expect that this calculation agrees well with angular differential data (since it manifestly does not at RHIC).

Thus, what is tested by this investigation is the response of parton-medium interaction model to a significant change in medium temperature (or equivalently density of scattering centers) and to the hardening of the primary parton production $p_T$ spectrum. While the first effect primarily governs an overall normalization factor of $R_{AA}$ since the strength of parton-medium interaction scales with the density of scatterers, the second effect leads to explicit $P_T$ dependence.

\begin{figure}[htb]
\epsfig{file=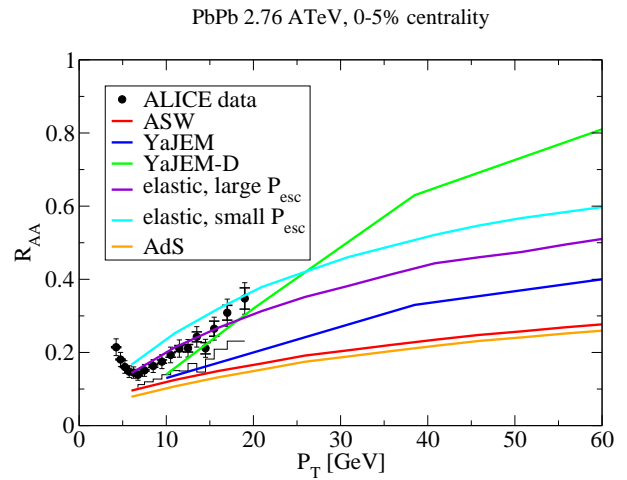, width=8cm}
\caption{\label{F-RAA-LHC} Various RHIC-constrained parton-medium interaction models (see text) extrapolated to 2.76 ATeV central Pb-Pb collisions at LHC using the EKRT minijet saturation model \cite{EKRT} to extrapolate a 2+1d hydrodynamical evolution along with a pQCD calculation for hard parton production \cite{RAA_LHC} compared with ALICE data \cite{ALICE-RAA}.}
\end{figure}

A comparison of various parton-medium interaction model when extrapolated from RHIC to LHC energies using the closed EKRT framework is shown in Fig.~\ref{F-RAA-LHC} (see \cite{RAA_LHC} for details of the calculation). It is evident that the $P_T$ dependence of various models indeed differs as expected, and also that the normalization of a suitable $P_T$ average is quite different, i.e. we observe both the response to the changed medium density and to the hardening of the parton spectrum.

It is clearly seen that both the ASW and the AdS frameworks (and to a lesser degree YaJEM) 'overquench', i.e. they predict more suppression than seen in the data. The overquenching of the ASW model appears to be generic for  radiative energy loss models, as it also appears for YaJEM and a qualitatively rather similar result was obtained within the WHDG model \cite{RAA_LHC_WHDG}. 

Given the overquenching of radiative energy loss models, an even stronger overestimation of the parton-medium interaction strength in a strong coupling description (AdS) is inevitable: As seen in \ref{SS-Interaction}, while parametrically the energy loss in radiative models scales as $\sim T^3 L^2$, in a strong coupling description this becomes $\sim T^4 L^3$. Since at higher $\sqrt{s}$ both mean temperature and (to a lesser degree) mean size of the medium increase, parametrically $R_{AA}$ in a strong coupling description \emph{must} extrapolate in $\sqrt{s}$ to a lower value than in a radiative description, thus if radiative models overquench, a strong coupling description will fare even worse.

With regard to the $P_T$ dependence, one can observe that it appears to be too weak in the majority of models, implying that the probability to loose a small amount of energy from the leading parton is underestimated by such models. The exception to this rules are parametrized elastic models with a Gaussian assumption for $P(\Delta E)$ \cite{elastic_phenomenology} which by construction leave more room for small energy losses as compared with the almost flat $P(\Delta E)$ characteristic for radiative energy loss models, and YaJEM-D which has an explicit energy dependence, i.e. can only be cast into the form $P(\Delta E|E)$. Since we already excluded a dominant contribution of an incoherent energy loss mechanism based on pathlength dependence in the previous section, the only relevant model at this point is YaJEM-D. In comparison with YaJEM, it can be observed that this explicit reduction of energy loss with increasing initial parton energy $E$ leads both to a different normalization and a steeper increase with $P_T$, in much better agreement with the data.

The same mechanism as in YaJEM-D (i.e. minimum scale down to which a shower is evolved in the medium determined by $Q_0 = \sqrt{E/L}$) has been used, albeit in a simplified implementation, in the context of the HT formalism \cite{RAA_LHC_HT}, and qualitatively in agreement with YaJEM-D, a strong increase of $R_{AA}$ with $P_T$ and no overquenching is observed.

We can infer from these observations that $P_T, \sqrt{s}$ dependence of $R_{AA}$ and centrality dependence provide in some sense orthogonal constraints. While the centrality dependence favours models in which energy loss scales with a high power of pathlength and rules out incoherent mechanisms, almost the opposite is true for the $P_T$ dependence at LHC where models with a weak pathlength dependence fare better with the data since they have a higher probability of inducing only small energy loss. 

The only class of models which can account for both observables simultameously introduces a minimum in-medium virtuality scale, as exemplified by both YaJEM-D \cite{YaJEM-D} and resummed HT \cite{HT-DGLAP,RAA_LHC_HT}. This implies that the dominant mechanism of leading parton energy loss is indeed likely to be medium-induced perturbatively calculable radiation, but this mechanism can not be treated in a leading parton energy loss model, since such a model can not implement a constraint on the minimum virtuality scale of a developing shower. Thus, coherence is an important ingredient, but on the level of the LPM effect as implemented in leading parton energy loss models it is effectively not much different from an incoherent mechanism \cite{Korinna-LPM,YaJEM-D,Caron-Huot}. Currently, there is no evidence that a strong coupling scenario is favoured (or even allowed) by the combined high $P_T$ observables.

\subsection{$z_T$ dependence of $I_{AA}$ at RHIC}

\label{S-IAA}

The suppression factor $I_{AA}$ of back-to-back correlated yield corresponds to a conditional probability of finding subleading hadrons on the near or away side given a trigger hadron in a certain momentum range. The requirement of a trigger biases the selection of events in various interesting ways as compared with a full jet observable where (in principle) every high $p_T$ scattering process contributes in an unbiased way \cite{Dihadron2}. With $z_T$ being the ratio of trigger momentum over associate hadron momentum, the low $z_T$ range of $I_{AA}$ then probes the fragmentation pattern of subleading shower hadrons given the biased selection of triggered events.

Due to the combination of biases, $I_{AA}$ is a fairly complicated observable, but it has several advantages over single inclusive hadron suppression. First, since the path of the away side parton through the medium is always different from the path of the near side parton, a simultaneous measurement of $R_{AA}$ and $I_{AA}$ allows to probe pathlength dependence of a parton-medium interaction model for central collisions, i.e.  \emph{without any change} in the hydrodynamical evolution from the point where model parameters are calibrated. Second, the near and away side correlated yield at low $z_T$ is sensitive to subleading fragmentation, i.e. probes the dynamics of energy redistribution within a shower beyond the validity of a leading parton energy loss model. Third, the trigger requirement leads (for RHIC kinematics more strongly than for LHC kinematics) to a bias for the away side parton to be a gluon and thus opens some sensitivity to the parton type.

The systematics of the dependence of $I_{AA}$ both on the medium evolution model and the parton-medium interaction model has been investigated in \cite{Dihadron2} and \cite{DihadronElastic}.

The main results can be summarized as follows: For $z_T > 0.5$ (i.e. for leading hadron physics), the systematics of $I_{AA}$ agrees with what can be obtained from $R_{AA}(\phi)$, i.e. the same pathlength dependences in combination with the same hydrodynamical models are preferred by both observables, although there are differences in the sensitivity in detail.

For $z_T < 0.5$, clear differences between models utilizing the energy loss approximation and full in-medium shower evolution codes are apparent. While energy loss models qualitatively show the wrong behaviour with the data, i.e. a decrease of $I_{AA}$ when approaching low $z_T$ from above, in-medium shower codes show an increase corresponding to the medium-induced additional gluon radiation. Note that in energy loss models this contribution is absent because the fate of subleading partons is not explicitly tracked by definition, and thus energy lost from the leading partons by construction disappears from all observables.

\begin{figure}[htb]
\epsfig{file=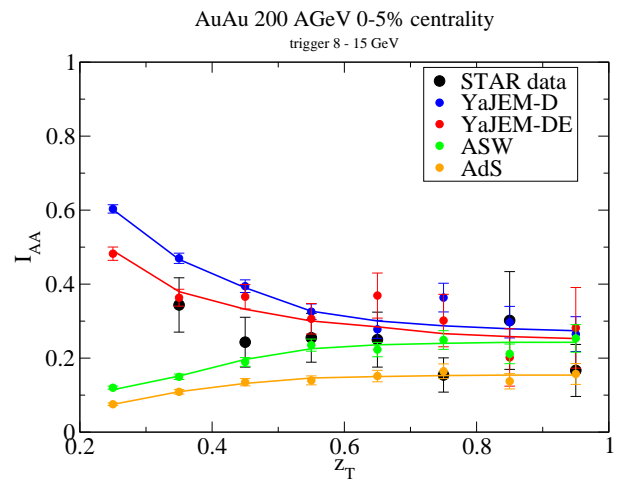, width=8cm}
\caption{\label{F-IAA} Away side dihadron suppression factor $I_{AA}$ computed with YaJEM in a pure radiative scenario (YaJEM-D) and with a 10\% contribution of elastic energy loss (YaJEM-DE) \cite{DihadronElastic} and for comparison also in the energy loss approximations models ASW and AdS using a 3+1 d hydrodynamical model for the medium \cite{hydro3d} compared with STAR data \cite{STAR-DzT}. Lines in theory results are drawn to guide the eye.}
\end{figure}

The observation of an upward trend of $I_{AA}$ for low $z_T$ in itself can thus be taken as evidence for a medium-induced radiation scenario, as this region probes the remnant of the radiation after hadronization explicitly. This is illustrated in Fig.~\ref{F-IAA} with results from YaJEM-D using the best choice for the medium model in comparison with model results using the energy loss approximation.  Clearly, the two energy loss models ASW and AdS miss the trend visible in the low $z_T$ data. As a side remark, note that the AdS model as shown in the figure overpredicts the amount of suppression seen in the data even at high $z_T$. This is driven by the choice of the hydrodynamical model, which is here selected to give the best fit using YaJEM-D. Based on such a figure alone, one might be tempted to rule out AdS as not viable with the data, however if a different medium model (\cite{hydro2d} for instance) would have been chosen, AdS would agree well with this set of data whereas YaJEM-D would not. This strongly emphasizes the point that the systematics associated with the medium modelling must be understood properly before drawing conclusions.

One can observe that a fully radiative scenario as exemplified by YaJEM-D overpredicts the rise at low $z_T$. However, any elastic contribution to the parton-medium interaction can be expected to deplete the medium-induced radiation spectrum by transfering energy from soft gluons into the medium as suggested e.g. in \cite{Collimation}. Allowing for a 10\% elastic contribution (i.e. about the maximal value that is allowed by pathlength-dependent observables) good agreement with the data is obtained \cite{DihadronElastic}. Note that such a contribution is small enough not to change any of the single hadron observables discussed above in a significant way, i.e. such a scenario (referred to as YaJEM-DE in the following) agrees with the results shown for YaJEM-D above.

We can infer from these results that the $z_T$ dependence of $I_{AA}$ confirms the constraints for pathlength and geometry as outlined above. In addition, the observation of the onset of the medium-induced radiation spectrum argues for a dominantly radiative energy loss picture and makes the treatment in terms of in-medium shower evolution mandatory --- energy loss modelling leads to even qualitative failure to describe the data at low $z_T$. The amount of medium-induced radiation is overpredicted by a pure radiative picture and hence allows to constrain the contribution of elastic interactions from below. This complements the constraints on elastic interactions from above by pathlength-sensitive observables, and both sets of constraints can be satisfied with a fraction of $\sim 10$\% elastic energy transfer. In principle, this value contains valuable information about the microscopic dynamics of parton-medium scattering (for instance, the amount of elastic energy transfer is a function of quasiparticle mass).

\subsection{Summary}

In summary, a combined analysis of models in the context of the measured $R_{AA}(P_T, \phi, \sqrt{s}$ and $I_{AA}(z_T)$, taking into account the systematic uncertainties associated with the medium evolution model suggests the following picture:

\begin{itemize}
\item The dominant fraction of the parton-medium interaction is coherent. A large incoherent contribution to the interaction necessarily leads to a linear parthlength dependence of energy loss, and this can be ruled out with great confidence due to the failure of linear pathlength models to describe the reaction plane dependence of $R_{AA}$. These findings are confirmed by the failure of linear pathlength dependence to account for $I_{AA}$ \cite{elastic_phenomenology}. 

\item Pathlength dependent observables rule out in addition to incoherent mechanisms also models in which coherence playes a role, but in which corrections such as finite kinematics or finite length effects effectively reduce the pathlength dependence to linear. Several theoretical results \cite{Korinna-LPM,YaJEM-D,Caron-Huot} suggest that this happens generically for radiative models implementing the LPM effect which generate an $L^2$ dependence of energy loss for infinite kinematics, i.e. the constraints from pathlength dependence are quite severe and affect a large class of suggested scenarios.

\item While $S^{in}_{out}$ tends to favour the scaling of energy loss with a high power of pathlength, dimensional arguments then suggest that energy loss then also scales with  correspondingly higher power of the medium temperature $T$. Such a scaling is not indicated by the observed $R_{AA}$ at the LHC --- radiative energy loss models parametrically scaling with $T^3$ tend to \emph{overpredict} the amount of suppression, correspondingly any model scaling energy loss $\sim L^n$ with $n>2$ will scale even worse as long as the average temperature in the LHC medium is higher than in the RHIC medium. This disfavours strong coupling models which can generate scaling with higher powers of $L$.

\item The combinaion of the above two points suggests that the pathlength dependence of energy loss is not just a simple power of $T$ as given e.g. by the LPM effect assuming infinite parent parton kinematics. This is also suggested by more detailed theoretical arguments. One scenario that is allowed by the data is a restriction on the minimum virtuality in the shower that can be probed in the medium as used by YaJEM-D or the resummed HT approach, other scenarios may be possible.

\item The observation of a rise in the low $z_T$ region of $I_{AA}$ is consistent with a radiative energy loss scenario. Qualitatively, additional medium-induced radiation is expected to produce such a rise. Quantitatively, a constrained calculation with a small fraction of elastic energy loss (YaJEM-DE \cite{DihadronElastic}) is able to account for the data. In this framework, the elastic fraction of energy loss can be constrained from above and from below to be about 10\%. The data at low $z_T$ cannot even qualitatively be explained in an energy loss picture, a full treatment of subleading shower partons is needed. 

\end{itemize}

If we always select the optimal medium evolution model for a given parton-medium interaction model, the findings of the systematic analysis can be summarized in table \ref{T-summary}.

\begin{table}
\begin{tabular}{|l|ccc|}
\hline
& $R_{AA}(\phi)$@RHIC & $R_{AA}$@LHC ($P_T$)& $I_{AA}$@RHIC  \\
\hline 
elastic &  {fails} &  {works} &  {fails} \\
elMC &  {fails} &  {fails} &  {fails}  \\
ASW &    {works} &  {fails} &  {marginal}  \\
AdS &    {works} &  {fails} &  {marginal}  \\
YaJEM &  {fails} &  {fails} &  {fails} \\
YaJEM-D &  {works} &  {works} &  {marginal}  \\
YaJEM-DE &  {works} &  {works} &  {works}  \\
\hline
\end{tabular}
\caption{\label{T-summary}Viability of different parton-medium interaction models tuned to the $P_T$ dependence of $R_{AA}$ in 200 AGeV Au-Au collisions given various data sets under the assumption that the best possible hydrodynamical evolution scenario is chosen. The various labels refer to: elastic \cite{elastic_phenomenology}, elMC \cite{elasticMC1}, ASW \cite{QuenchingWeights}, AdS \cite{hybrid3}, YaJEM \cite{YaJEM1,YaJEM2}, YaJEM-D \cite{YaJEM-D}, YaJEM-DE \cite{DihadronElastic}.}
\end{table}

It becomes readily apparent that some constrains work in an orthogonal way, for instance while $R_{AA}$ at LHC does not lead to conclusive statements about elastic energy loss scenarios, any pathlength dependent observable quickly establishes their failure.

Of the models for which the full systematics has been tested, only one combination (YaJEM-DE \cite{DihadronElastic} in the 3+1d ideal hydrodynamical code by Nonaka and Bass \cite{hydro3d} results in a viable description of all the combined data discussed so far. Other models (for instance the HT approach) for which the full systematics with realistic medium models is not known exhibit the properties required by the data, and there is no reason to assume such models would not lead to an equally good description of the data. Given that the full matrix of seven \emph{a priori} viable parton-medium interaction models (based on a description of $R_{AA}(P_T)$ in 200 AGeV central Au-Au collisions at RHIC) combined with four possible choices of the medium evolution model for which the systematics is known contains 28 different combinations, the constraining power of the multi-observable analysis becomes readily apparent.

\section{Open issues}

\label{S-OpenIssues}

This section summarizes various observables or model frameworks which have, for reasons given for each in the specific context, not been included into the analysis of the previous section.

\subsection{$R_{AA}$ of protons}

As discussed in \ref{SS-Fragmentation} one of the central assumptions underlying the theory of hard probes is that the final state interaction is predominantly a partonic phenomenon, i.e. that the observed effect is not driven by hadron-hadron interactions during the later stage of the medium evolution. The obvious way to test this assumption is to study observables for various identified hadron species.

If the final state interaction is predominantly a partonic phenomenon, then $R_{AA}$ for various hadron species should depend on the relevant partonic production channels for the hadron, but not on the actual identity of the hadron. Within the AKK set of fragmentation functions \cite{AKK} the proton is found to be dominated by the fragmentation of gluons in the RHIC kinematic range 5-10 GeV whereas the pion receives about equal contribution from quark and gluon fragmentation. Since gluons interact more strongly with the medium by a color factor $C_F = 9/4$, the natural expectation is that proton $R_{AA}^{p}(P_T) < R_{AA}^{\pi}(P_T)$. Experimentally however, the opposite trend was observed \cite{STAR-RAA-pp}.

On face value, this finding argues against a partonic origin of the suppression and it is currently not well explained. In a detailed computation using the ASW formalism, the nuclear suppression factor of pions and protons is found to be fairly similar, but as expected with $R_{AA}^{p}(P_T) < R_{AA}^{\pi}(P_T)$ \cite{RAA-pp}. If conversion reactions like $q\overline{q} \rightarrow gg$ which change the identity of the hard parton are introduced \cite{Conversion} then proton and pion $R_{AA}$ can at best be made equal. It should be noted however that one of the central assumptions made in almost every model of parton-medium interaction, i.e. that the hadronization process takes place outside the medium, is violated in the kinematic range of the data. If the formation time of protons is estimated as $\tau \sim E_p/m_p^2$ a value of $\sim 2$ fm is obtained, i.e. it can not be argued that proton formation happens outside the medium, instead the scale indicates that in most cases it happens inside the medium. Thus, models which assume a vacuum hadronization pattern are not applicable to this data.

\subsection{Heavy quark energy loss}

The idea to study the energy loss of heavy $c$ and $b$ quarks is based on the notion that their radiative energy loss contribution should be significantly reduced as compared to light quarks due to the so-called dead cone effect \cite{DeadCone}. This has led to the expectation that heavy quark $R_{AA}$ (as probed by single inclusive  electron suppression) should be observed above light quark driven hadronic $R_{AA}$. However, the single inclusive non-photonic electron suppression measurements (which predominantly probes heavy quark decays) have shown a comparable magnitude of heavy and light quark suppression \cite{HQexp1,HQexp2}. This has sometimes been refered to as 'heavy quark puzzle'  and lead to the conjecture of a large elastic energy loss component $\sim 50$\% \cite{HQPuzzle}. 

Such a large component in the light quark sector is however, as we have seen in previous sections, conclusively ruled out by pathlength-sensitive observables. Thus, one is forced to either conclude that a large elastic component is not the solution to the heavy quark puzzle or that the physics of light and heavy quark energy loss is significantly different. There is in fact some reason to suspect the latter scenario.

First, the energy loss due to elastic collisions is expected to be enhanced relative to light quarks due to the 'tagging' effect \cite{HQEloss} --- while for an unidentified primary quark a collision leading to a transfer of 80\% of its energy to a medium parton effectively counts as 20\% energy loss since the medium parton then becomes the new hard parton, a heavy quark is always identified since due to the high mass threshold it cannot be generated by thermal excitation and hence 80\% energy transfer in this case effectively imply 80\% energy loss.

A second important aspect which is not obvious in an energy loss approximation becomes apparent in a medium-modified shower picture. The formation time of a heavy-quark induced shower, i.e. the time it takes to reduce the original quark virtuality at the hard scale to the quark mass scale, and also the formation time of heavy mesons is very short \cite{HQdissociation}. This would imply that the object interacting with the medium is in fact not a quark, but rather a pre-hadronic resonance \cite{HQdissociation,HQdiffusion} which is quickly dissociated by the medium, but nevertheless enhances the effective interaction cross section.

Models based on elastic heavy-quark medium interaction can explain the existing data \cite{HQdissociation,HQdiffusion,HQTomography1,HQTomography2}. An alternative approach is to postulate the same physics for light and heavy quark energy loss and treat heavy quark energy loss using the strong coupling assumption \cite{HQAdS1}. As we have argued above, this is not in agreement with the scaling hadronic $R_{AA}$ from RHIC to LHC energies in the light quark sector. As pointed out e.g. in \cite{HQAdS2}, expectations for heavy quark $R_{AA}$ at LHC are substantially different dependent on if a perturbative or strong coupling scenario is used.

In view of the fact that both the tagging and the formation time argument imply that the physics of heavy quark energy loss is different beyond a kinematic dead cone radiation suppression and may quite possibly contain non-perturbative QCD dynamics, the non-photonic single electron data has not been used for the analysis presented in this work.

\subsection{$\gamma$-hadron correlations}

In principle, $\gamma$-h correlations allow to constrain the energy of the recoiling parton at the hard vertex much better than dihadron correlations. In practice, the reduced experimental statistics often requires a comparatively wide trigger momentum window which tends to offset this advantage somewhat. While $\gamma$-h correlations probe the fragmentation function in the limit where the photon energy is known with great accuracy, the opposite situation in which photons within a wide energy range are accepted as triggers resembles more an $R_{AA}$ of quarks since the $gq \rightarrow \gamma q$ production channel dominates over $q\overline{q} \rightarrow \gamma g$ and since, unlike for dihadron correlations, the photon does not bias the distribution of initial vertices given a trigger. 

From these considerations, it can be deduced that any parton-medium interaction model which is tuned to describe single hadron suppression for a given choice of a medium model should also describe the high $z_T$ region of the $\gamma$-h data \cite{gamma-h-STAR,gamma-h-PHENIX} in the same medium model because the integrals over geometry and energy loss probability involved are just the same. This is indeed found for the Zhang-Owens-Wang-Wang (ZOWW) model \cite{gamma-h-Zhang}, for the AMY model \cite{gamma-h-Qin}, for the ASW model and for the shower code YaJEM \cite{gamma-h-YaJEM}.

The interesting differences between the models (and between model and data) appear at low $z_T$. For instance, as in the case of dihadron $I_{AA}$, YaJEM is found to overshoot $\gamma$-h $I_{AA}$ at low $z_T$ whereas ASW does not show any rise coming from high $z_T$ down to low $z_T$. This suggests the same basic physics interpretation: The rise at low $z_T$ measures the strength with which medium-induced radiation leads to subleading shower hadrons. A computation of this contribution requires a full in-medium shower framework, and would constrain the contribution of elastic energy loss from below.

At present, it is unclear if there is information in $\gamma$-h correlation that can not be obtained from the combination of hadronic $R_{AA}$ and dihadron $I_{AA}(z_T)$. For this reason and the lack of published systematic studies with different medium models, the observable has not been used for the analysis presented in this work. However, unlike in the case of proton $R_{AA}$ or heavy quark energy loss, there is currently no reason to assume that describing $\gamma$-h correlations would be problematic for any model which fits the known constraints.

\subsection{The dijet asymmetry}

The energy imbalance between back-to-back jets in a medium has been quantified by ATLAS and CMS \cite{ATLAS,CMS} and has immediately sparked a lot of theoretical activity.  Models able to account for the measured imbalance range from parametric estimates \cite{Collimation} via schematic modelling of energy loss and jet finding \cite{Dijets-Qin} to sophisticated NLO pQCD modelling of jet evolution in a medium, however without detailed medium or jet-finding modelling \cite{Dijets-Vitev} and complete MC modelling of a jet embedded into a bulk medium with realistic jet finding using the MARTINI code \cite{Dijets-Martini}.

At this point, this clearly argues that there is nothing mysterious about the measured dijet imbalance which would require an exotic scenario of parton-medium interaction and that  pQCD based models are quite able to account for the data. Given the lack of systematic studies with realistic medium geometries and the uncertainties still associated with a correct determination of parton energy given a jet in a medium background, the observable has not been included into the analysis presented here.

There is currently no reason to expect that for instance YaJEM-DE as constrained from $R_{AA}$ and $I_{AA}$ would not be in agreement with the dijet asymmetry data. Qualitatively, the YaJEM-DE is very similar the scenario described in \cite{Collimation}, i.e. medium-induced soft gluon radiation is depleted quickly by elastic processes whereas hard gluon radiation is more robust and survives as part of the jet. From the results of  \cite{YaJEM-Jets} it is expected that medium-modified jets largely resemble jets in vacuum but shifted in energy.  However, a (numerically rather involved) quantitatve comparison is still missing.

\subsection{Color (de-)coherence in the medium}

An essential ingredient to the detailed description of a developing parton shower in vacuum is color coherence, leading effectively to the phenomenon of angular ordering, i.e. the angles between parent and daughter parton in subsequent gluon emissions decrease. For in-medium radiation on the other hand, it has been found that anti-angular ordering is obtained \cite{AntiAngularOrdering} and that decoherence of the total radiation spectrum in a sufficiently dense medium follows \cite{Decoherence}.

Such a density-driven transition from color coherence to decoherence has so far not been included into any model of parton-medium interaction, thus all models currently being compared to data miss a potentially important contribution to the dynamics.

There are however two reasons to believe that this is not a huge problem in practice. First, if one estimates the formation time by the uncertainty relation as parametrically $\tau \sim E/Q^2$ where $E$ is the parton energy and $Q^2$ the current virtuality scale, then the first hard branchings (which largely determine the structure of a jet in terms of e.g. subjet structure or transverse broadening) happen at timescales much smaller than any medium formation timescale as discussed in e.g. hydrodynamical initial states.  This implies that whatever a dense medium will do to color coherence afterwards, the basic structure of a jet is determined by color coherence just as in vacuum since the medium cannot yet have formed. If a medium could be prepared beforehand and a jet embetted into this medium, then color decoherence would be a much more severe issue, but this is a different problem which does not correspond to the experimental situation.

Second, while the transition to color decoherence is a difficult problem, the difference between full color coherence and full decoherence can be explored in in-medium MC shower models. This has been done e.g. in \cite{YaJEM2} and was found to be a small effect as compared to other model uncertainties. Thus, while color decoherence is a very interesting theoretical problem which is not sufficiently addressed in current modelling and highly relevant for precision jet physics, there is currently no reason to assume that it would be a major effect in practice or that it would invalidate any of the results given above.

\subsection{Core-corona picture}

In order to explain the large $S^{in}_{out}$ observed at RHIC kinematics, core-corona models like \cite{CoronaPantuev} have been postulated in which the parton-medium interaction is essentially geometry driven: If, after a certain formation time $\tau_{form}$, the parton is still in the interaction region, the parton is absorbed by the medium, otherwise the parton escapes without substantial interaction.

In the context of angular dependent observables, such geometrical models work in general better than pQCD energy loss computations. However, they fail for other observables. In essence, a formation time in the energy loss picture can be written as a pathlength $L$ dependence like $\Delta E \sim \left( \frac{L}{c \tau_{form}} \right)^n$ with $n$ a large number. Such a model would thus be disfavoured by the LHC data in the same way as the AdS $L^3$ dependence of energy loss, i.e. a purely geometric model could not generate sufficient $P_T$ dependence to account for the data.

Another problem is the actual parameter $\tau_{form}$ needed to account for the data. In \cite{CoronaPantuev} this is found as $\sim 2.3$ fm, i.e. significantly longer than the formation times needed by typical hydrodynamical bulk models. While it is pointed out correctly that the hydrodynamical formation time $\tau_i$ measures the time by which the medium is sufficiently equilibrated and pressure gradients act, and that this doesn't need to be equal to the time at which parton-medium interaction is strong, the difference unfortunately acts in the wrong way: Parton-medium interactions probe the density of color charges, and color charges must be present in the system before it can equilibrate, thus $\tau_{form} < \tau_i$ is the expected ordering.  

The observed form of $I_{AA}(z_T)$ as discussed in \ref{S-IAA} likewise argues against a geometrical picture in the sense that the enhancement seen at low $z_T$ is qualitatively and quantitatively consistent with hadronizing medium-induced radiation. In a core-corona picture where a parton is either absorbed or escapes unmodified, no reason for such an enhancement would be present.

Thus, while a geometrical interpretation is tempting (and indeed superior) given a subset of the high $P_T$ data, it is insufficient to account even qualitatively for other parts of the data and has for this reason not been considered in the main analysis section of this work.

\section{Conclusions}

A systematic study of combinations of parton-medium interaction models and bulk medium fluid dynamical evolution models against a large body of available precision data on high $P_T$ observables allows to establish the following points:

\begin{itemize}

\item The basic strategy works, i.e. the multi-observable analysis has constraining power for both parton-medium interaction model and bulk medium evolution model. The constraints posed by the data are significant, more than 95\% of the scenarios tested showed a failure to describe the data for at least one observable. However, the strategy relies strongly also on the use of hydrodynamical modelling which in itself must be constrained against bulk observables, it would fail if also parametrically plausible medium model choices without explicit constraints from the data (Bjorken cylinder, hard sphere overlap...) would be acceptable. 

\item The constraints on the parton-medium interaction side identify a number of features of the relevant physics process: The dominant energy transfer mechanism from the hard parton to the medium must be coherent. It leads to induced radiation which is observed in experiment and is qualitatively in agreement with medium induced gluon radiation in a pQCD scenario. A subdominant, elastic component is needed to deplete and decorrelate the soft gluon spectrum from the jet structure via energy and momentum transfer into the bulk medium, this component can be constrained from above and from below to about 10\%. Presumably, there is an explicit dependence on the initial hard parton energy in the MMFF, leading to $P(\Delta E, E)$ rather than $P(\Delta E)$ in the energy loss approximation. It follows from the above that modelling finite kinematics and full in-medium shower development is crucial to get agreement with all features of the data.

\item Constraints for the hydrodynamical side favour generally a late onset of energy loss and a large extension of the freeze-out hypersurface in space. Both a steeper initial density profile as found in CGC scenarios and viscous entropy generation are favoured by the data, but somewhat surprisingly are not dominant effects. In spite of commonly made assumptions, there is no reason to identify the equilibration time of the medium $\tau_0$ or the freeze-out isothermal surface characterized by $T_F$ as the relevant scales for the parton-medium interaction. 

\item Other high $P_T$ observables so far do not substantially challenge this picture: Both heavy quark suppression and identified proton suppression are known to contain different and non-perturbative physics, as the hadronization cannot be argued to take place outside the medium, thus it is likely that the apparent puzzles associated with these observables translate into constraints for this new ingredient. Jet observables at the LHC are qualitatively in agreement with the picture outlined above, quantitative comparisons are difficult and numerically challenging due to various jet-specific issues, but in many cases remain to be done.

\end{itemize}

The weakest point of almost all models studied here remain observables which are differential in the reaction plane angle $\phi$. It is in general non-trivial to reproduce the observed magnitude of $S^{in}_{out}$ --- even the calculations which are tuned to maximize this quantity barely reach above the measured value.  Most probably this has to be be attributed to a problem on the hydrodynamical side rather than the parton-medium interaction side, since $I_{AA}$ for instance also probes pathlength dependence of the medium effect, but shows no problem comparable to $S^{in}_{out}$. This needs further study including initial state fluctuations of the bulk medium and different choices for the exit criterion of a parton from the medium. 

Since the strategy used in this work exploits qualitative differences to discriminate between various models, at this point no conclusion with regard to numerical values of medium properties such as the transport coefficient $\hat{q}$ can be drawn. This is largely due to what has been summarized before as 'implementation details' and discussed in depth in \cite{Brick}: Internal model parameters such as cutoffs lead to a large uncertainty in the extraction of numerical values. It is unlikely that all this uncertainty can be overcome by a data driven approach, thus improved theoretical calculations based on models which contain the right qualitative dynamics is needed to turn hard probes into a quantitative measurement of medium properties.

\begin{acknowledgements}
 This work was 
supported by an Academy Research Fellowship of T.R. from the Finnish Academy (Project 130472) and 
from Academy Project 133005. Discussions with Kari Eskola, Jussi Auvinen, William Horowitz, Bj\"{o}rn Schenke, Hannu Holopainen, Ulrich Heinz, Abhijit Majumder, Jan Rak, DongJo Kim, Andreas Morsch, Marco van Leeuwen, Peter Jacobs, Megan Connors, Ahmed Hamed, Andre Mischke, Barbara Jacak, Jan Fiete Grosse-Oetringhaus, Andrew Adare and Risto Paatelainen are gratefully acknowledged.

\end{acknowledgements}

\end{document}